# STUDY OF THE HII REGIONS IN THE SPIRAL GALAXY NGC6384

A.A.HAKOBYAN[1], A.R.PETROSIAN[1], A.A.YEGHIAZARYAN[1], J.BOULESTEIX[2]

The galaxy NGC6384 has been observed with an IPCS through Hα and [NII] narrow-band interference filters for direct imagery with the 2.6 m Byurakan telescope. We studied main physical parameters of identified 98 HII regions, their diameter and luminosity functions, as well [NII]/Hα ratio distribution. The integrated distribution function of the HII region diameters can be well fitted by the exponential function. The characteristic diameter has the value ($D_o$ = 217 pc) predicted for a galaxy of its measured absolute luminosity. The luminosity function of HII regions has double power laws profile with relatively shallow slope at low luminosities ($\alpha$ = - 0.4), an abrupt turnover at log $L$(Hα) ≈ 38.75, and sharper slope at higher luminosities ($\alpha$ = - 2.3). Correlation between luminosity and diameter of HII regions confirms that in general they are constant density, radiation-bound systems. [NII]/Hα ratio data for HII regions show that there is a negative radial gradient of [NII]/Hα. In the central region of the galaxy, nitrogen abundance is higher than in the periphery. The properties of the HII regions population of this AGN galaxy does not differs significantly from the properties of the HII regions population of the "normal" galaxies. Reexamining the location of the type Ia SN 1971L in the galaxy, we confirm that it lies on the spiral arm at about 8.6″ far from the closest HII region #53 (F81). Such a location can be taken as prove that the progenitor of this SN do not belong to an old, evolved stellar population.

1. *Introduction.* The relationship between the characteristics of the active nucleus and the global properties of the host galaxies is of particular interest in understanding the nature of the active galactic nuclei (AGNs) phenomenon. Available data on the AGN-host galaxy connection are rare and often contradictory. Largest study in this field presented by Kauffmann et al. in [1] concluded that the host galaxy properties depends form AGN activity level, with high [OIII] luminosity AGNs hosts have much younger mean stellar ages. This result is based on the statistical study of more than 22600 SDSS galaxies and will be valuable to prove it by detail study of the individual objects. Star formation in the disk of spiral galaxies is usually marked by the presence of HII regions that are thought to be the sites of young stars. HII region diameter and luminosity functions and their location in the galactic disk can potentially add to our understanding of the star formation process in galaxies. Differential investigation of the HII region populations in normal and active spiral galaxies may help us reveal, in details, possible differences between the star-forming ability of the host galaxies with AGN or normal nucleus.

In this paper we study in details associated with stellar associations or super assosiations HII region population of the galaxy NGC6384 with LINER(T2) type AGN (e.g. Hughes et al. [2]). This galaxy is one of the best studied near by (D = 22.2 Mpc; calculated according to the recent $v_o$ = 1665 km/s value and $H_o$ = 75 km $s^{-1}$ $Mpc^{-1}$), SB(r)bc type (e.g. Eskridge et al. [3]) spiral galaxies with low inclination ($i$ = 47°) angle and well developed spiral arms (e.g. Garsia-Gomez et al. [4]). Because of low inclination of this galaxy its HII region population is well studied. At first Hodge and Kennicutt [5] identified and mapped 142 HII regions of NGC6384. Feinstein [6] identified 283 HII regions in this galaxy and also measure their Hα luminosities. Recently Knapen et al. [7]

presented calibrated high resolution Hα map of NGC6384. HII region distribution of this galaxy is studied in several articles (e.g. Athanassoula et al. [8]). The results of optical and near-infrared surface photometry of NGC6384 are presented in several other studies (e.g. Elmegreen and Elmegreen [9], Sanchez-Portal et al. [10]). In the past years chemical abundance of 16 bright HII regions of NGC6384 have bas been measured (e.g. Bresolin and Kennicutt [11], Bresolin et al. [12]) and results are summarized by Pilyugin et al. [13].

In the following sections 2 and 3 of this paper we present the observations and reduction techniques and measurements of HII region diameters, positions, their integral Hα luminosities and [NII]/Hα ratios. Results on diameter distribution, luminosity function, spatial and [NII]/Hα ratio distribution of HII regions is presented in the section 4. Conclusions are drawn in the last section 5. Throughout this paper a value of the Hubble constant of $H_o$ = 75 km $s^{-1}$ $Mpc^{-1}$ will be assumed.

2. *Observations and data reduction.* The observations have been made with the CIGAL equipment mainly consisting of a focal reducer with IPCS and scanning Fabry-Perot interferometer (Amram et al. [14], Marcelin et al. [15]) attached to the *F/3.8* Prime focus of the 2.6 m. telescope of Byurakan Observatory (Armenia) on September 7, 1985. An automated focal reducer gave a final aperture ratio of *F/1.9* and a total field of view 7.3′ x 7.3′ on the 256 x 256 pixels of the photon counting camera. The spatial scale is 1.7″ $pixels^{-1}$ on the detector. The CIGAL equipment operates in two direct imaging and interferometric modes. We used only direct imaging mode and imaging of NGC6384 was achieved through narrow-band interference filters; one centered on the redshifted Hα and another on redshifted [NII]λ6584Å emission lines. When corrected for the temperature and the focal ratio of the telescope, the filters had the following characteristics, for Hα central wavelength 6600.2Å and FWHM 10Å, and for [NII] line central wavelength 6620.2Å and FWHM 10Å. In both cases exposure time was 30 minutes. Both images were obtained during the same night using the same guiding star, so the light of the observed HII regions of NGC6384 fell on the same pixels for each image. Star-free sky regions were measured around the galaxy and subtracted from the Hα and [NII] images.

The identification and measurement of the HII regions have been done separately in Hα and [NII] line frames. Only HII regions for which the signal to near background ratio is more that three (3σ level) both in Hα and [NII] lines were retained. Because of overlap due to proximity of objects to each other and blurring by the seeing, two or more HII regions can seem to be one. To avoid such problem a careful inspection of each HII region was made. Using the criterion that each HII regions has a single peak of its center, objects with multiple peaks were assumed to consist of several independent objects, and attempt was made to identify the individual parameters. For identified HII regions their positions, diameters and [NII]/Hα ratios were calculated. The Hα equivalent diameters of the identified HII regions have been obtained in terms of the limiting isophotal area, *A*, according to the definition $D = 2(A/\pi)^{1/2}$. The limiting Hα isophote of HII regions is placed at a 3σ level above the local background. Apparent diameters of HII regions are enlarged by the instrument and by the seeing. The true diameter is given by: $D_{true} = (D_{obs}^2 - D_{psf}^2)^{1/2}$ where $D_{true}$ is the true diameter, $D_{obs}$ is the observed diameter and $D_{psf}$ is the instrumental profile or the point spread function, which has a FWHM of 1.7″ as derived from stellar images in the field. We also determine Hα luminosities of HII regions. These luminosities were calculated according to Feinstein [6] Hα + [NII] luminosities, which were corrected for the [NII] lines luminosities contribution using measured [NII]/Hα ratios. According to the Feinstein [6] the flux of an HII region is the integral of the signal inside of our adopted limiting isophotal area. [NII]/Hα ratios are calibrated using spectroscopic data for 9 HII regions (Oey and Kennicutt [16], Bresolin et al. [12], Bresolin and Kennicutt [11]).

3. *HII regions.* In total, on both Hα and [NII] images 98 HII regions have been identified. In Hα line the 142 HII regions of NGC6384 have been catalogued by Hodge and Kennicutt [5] and 283 by Feinstein [6]. Using measured positions of discovered HII regions we identified them with appropriate counterparts in [5] and [6] catalogues.

Table 1 lists the parameters of the HII regions measured in NGC6384. Column 1 gives our identification numbers of HII regions, and columns 2 and 3 present their RA and Dec coordinates in arcsecs. The HII regions positions were obtained as the pixel coordinates of maximum flux and these were referred to the center of the galaxy determined from the pixel of maximum flux in the Hα image. For RA and Dec eastward and northward values respectively are positive. Columns 4 and 5 give the appropriate numbers of identified HII regions according respectively to Hodge and Kennicutt [5] and Feinstein [6]. For NGC6384 assuming $i = 47°$ and $PA = 40°$ (Garsia-Gomez et al. [4]) column 6 presents, galactocentric deprojected distances of HII regions in units of the R(0) = 202.8 arcsecs radius of the galaxy (RC3, De Vaucouleurs et al. [17]). Column 7 gives the Hα effective linear diameters (*D*) of the identified HII regions. Column 8 presents logarithm of Hα luminosities and column 9 gives their [NII]λ6584 to Hα lines ([NII]/Hα) ratios. In Fig.1 we show schematically the positions of HII regions in the disc of NGC6384, on a RA-Dec grid centered on the nucleus of the galaxy.

*Table 1*

PARAMETERS OF THE HII REGIONS MEASURED IN NGC6384

| HII | RA (arcsec) | Dec (arcsec) | HK(1983) | F(1997) | R/R(0) | *D* (pc) | log *L*(Hα) | [NII]/Hα |
|---|---|---|---|---|---|---|---|---|
| 1 | 2 | 3 | 4 | 5 | 6 | 7 | 8 | 9 |
| 1 | 144.53 | 122.18 | 1 | 52 | 0.952 | 289 | 38.85 | 0.25 |
| 2 | 143.74 | 112.10 | 2 | 10 | 0.924 | 780 | 39.29 | 0.25 |
| 3 | 130.87 | 185.03 | | 118 | 1.126 | 392 | 38.43 | 0.26 |
| 4 | 122.31 | 39.39 | | 19 | 0.732 | 720 | 39.07 | 0.30 |
| 5 | 121.52 | 78.94 | 6 | 219 | 0.750 | 189 | 37.81 | 0.31 |
| 6 | 108.77 | 135.67 | 12 | 25 | 0.860 | 728 | 39.03 | 0.25 |
| 7 | 104.49 | -12.99 | 13 | 76 | 0.699 | 355 | 38.68 | 0.30 |
| 8 | 99.53 | 61.07 | 14 | 78 | 0.609 | 332 | 38.69 | 0.27 |
| 9 | 94.54 | 24.27 | 17 | 71 | 0.567 | 436 | 38.73 | 0.29 |
| 10 | 92.79 | 104.62 | | 96 | 0.691 | 438 | 38.58 | 0.29 |
| 11 | 91.14 | 46.65 | | 187 | 0.548 | 199 | 38.00 | 0.31 |
| 12 | 91.14 | 12.37 | 19 | 92 | 0.560 | 421 | 38.59 | 0.30 |
| 13 | 90.86 | 120.31 | 22 | 12 | 0.747 | 911 | 39.25 | 0.26 |
| 14 | 89.05 | 2.23 | 20 | 45 | 0.565 | 386 | 38.89 | 0.30 |
| 15 | 84.86 | 4.83 | 25 | 88 | 0.533 | 265 | 38.63 | 0.30 |
| 16 | 84.57 | -44.50 | | 145 | 0.683 | 210 | 38.25 | 0.29 |
| 17 | 83.27 | 38.01 | 21 | 134 | 0.497 | 297 | 38.32 | 0.28 |
| 18 | 81.29 | 132.84 | | 198 | 0.780 | 192 | 37.93 | 0.32 |
| 19 | 79.08 | 43.98 | 30 | 85 | 0.479 | 300 | 38.66 | 0.27 |
| 20 | 73.24 | -30.82 | 28 | 37 | 0.561 | 508 | 38.95 | 0.30 |
| 21 | 71.31 | -68.22 | 33, 34 | 38, 72 | 0.714 | 534 | 39.14 | 0.30 |
| 22 | 70.35 | 136.75 | | 23, 150 | 0.782 | 658 | 39.10 | 0.24 |
| 23 | 67.91 | 120.26 | 36 | 62 | 0.696 | 613 | 38.78 | 0.25 |
| 24 | 65.41 | 33.96 | 29 | 61 | 0.394 | 347 | 38.76 | 0.30 |
| 25 | 62.93 | -76.72 | 35 | 29, 57 | 0.713 | 511 | 39.18 | 0.33 |
| 26 | 62.87 | 28.63 | 37 | 35 | 0.376 | 248 | 38.96 | 0.30 |
| 27 | 59.58 | 69.43 | 43 | 32 | 0.452 | 502 | 38.97 | 0.29 |
| 28 | 59.47 | 32.14 | 42 | 31 | 0.359 | 299 | 38.97 | 0.30 |
| 29 | 58.73 | -43.45 | 38 | 109 | 0.528 | 447 | 38.48 | 0.29 |
| 30 | 57.83 | 43.99 | 44 | 15 | 0.369 | 373 | 39.20 | 0.31 |
| 31 | 57.83 | 15.08 | | 197 | 0.347 | 209 | 37.93 | 0.31 |
| 32 | 53.80 | 37.07 | 46 | 21 | 0.335 | 353 | 39.05 | 0.29 |
| 33 | 50.91 | 22.90 | | 105 | 0.303 | 296 | 38.49 | 0.31 |
| 34 | 50.40 | 61.09 | 48 | 126 | 0.392 | 299 | 38.35 | 0.29 |
| 35 | 48.47 | 10.73 | 47 | 91 | 0.293 | 267 | 38.59 | 0.31 |
| 36 | 48.36 | -4.30 | | 99 | 0.319 | 347 | 38.53 | 0.33 |
| 37 | 45.02 | 54.87 | 51 | 112 | 0.351 | 431 | 38.47 | 0.28 |
| 38 | 44.23 | 1.48 | 49 | 66 | 0.278 | 254 | 38.75 | 0.31 |
| 39 | 42.30 | -6.17 | 53 | 51 | 0.285 | 299 | 38.82 | 0.32 |
| 40 | 39.18 | 25.34 | | 56 | 0.241 | 471 | 38.80 | 0.29 |
| 41 | 39.13 | -67.88 | | 177 | 0.549 | 204 | 38.06 | 0.33 |
| 42 | 37.48 | 118.79 | | 87 | 0.667 | 473 | 38.65 | 0.29 |
| 43 | 37.37 | 81.38 | | 144 | 0.461 | 235 | 38.25 | 0.30 |
| 44 | 36.42 | 18.20 | | 46 | 0.218 | 299 | 38.88 | 0.30 |
| 45 | 35.73 | -0.22 | 54 | 80 | 0.227 | 248 | 38.67 | 0.31 |
| 46 | 33.91 | -29.80 | 55 | 28 | 0.326 | 298 | 38.96 | 0.34 |
| 47 | 31.53 | 101.67 | 57 | 50, 94 | 0.571 | 689 | 39.04 | 0.28 |
| 48 | 31.37 | -12.97 | 56 | 24 | 0.239 | 292 | 38.99 | 0.34 |
| 49 | 29.78 | 48.29 | 58 | 47 | 0.285 | 447 | 38.89 | 0.29 |
| 50 | 27.97 | -18.91 | 61 | 18 | 0.242 | 299 | 39.09 | 0.31 |
| 51 | 27.23 | 111.87 | 59, 63 | 17, 63 | 0.632 | 917 | 39.29 | 0.28 |
| 52 | 23.20 | -68.33 | 60 | 3 | 0.483 | 561 | 39.35 | 0.35 |
| 53 | 21.96 | 26.87 | | 81 | 0.172 | 248 | 38.68 | 0.28 |
| 54 | 16.29 | -74.05 | 64 | 163 | 0.491 | 241 | 38.14 | 0.31 |
| 55 | 15.27 | -10.42 | | 161 | 0.131 | 208 | 38.16 | 0.32 |
| 56 | 12.95 | 63.36 | 65, 70 | 55, 164 | 0.361 | 785 | 38.89 | 0.29 |
| 57 | 12.83 | 28.68 | | 7 | 0.163 | 465 | 39.31 | 0.30 |
| 58 | 6.49 | -26.00 | 68 | 4 | 0.173 | 534 | 39.35 | 0.33 |
| 59 | 5.01 | 37.24 | 76 | 13 | 0.215 | 658 | 39.22 | 0.30 |
| 60 | 0.20 | 51.69 | | 120 | 0.311 | 436 | 38.40 | 0.29 |
| 61 | -2.66 | -60.39 | 78 | 67 | 0.356 | 417 | 38.74 | 0.32 |
| 62 | -3.27 | -85.50 | 73, 77, 80 | 1, 33, 39 | 0.505 | 983 | 39.75 | 0.38 |
| 63 | -3.32 | -41.81 | | 9 | 0.243 | 338 | 39.26 | 0.33 |
| 64 | -3.50 | 123.88 | | 228 | 0.755 | 197 | 37.76 | 0.30 |
| 65 | -4.22 | 29.64 | 81 | 34 | 0.191 | 413 | 38.95 | 0.33 |
| 66 | -11.93 | -29.84 | 85 | 8 | 0.167 | 375 | 39.27 | 0.32 |
| 67 | -16.97 | 84.73 | 91 | 43 | 0.559 | 508 | 38.89 | 0.30 |
| 68 | -18.73 | -45.37 | | 227 | 0.254 | 204 | 37.77 | 0.29 |
| 69 | -20.03 | -8.14 | | 64 | 0.120 | 324 | 38.75 | 0.30 |
| 70 | -22.02 | 30.38 | 96 | 22 | 0.270 | 341 | 39.02 | 0.32 |
| 71 | -22.24 | 9.58 | 97 | 16 | 0.174 | 394 | 39.13 | 0.31 |
| 72 | -23.09 | 76.28 | | 60 | 0.532 | 248 | 38.76 | 0.33 |
| 73 | -25.53 | 1.60 | 99 | 26 | 0.169 | 320 | 38.99 | 0.33 |
| 74 | -27.17 | -22.32 | | 125 | 0.178 | 230 | 38.37 | 0.29 |
| 75 | -29.72 | -79.32 | 102 | 5 | 0.444 | 436 | 39.33 | 0.31 |
| 76 | -30.58 | 77.08 | 103 | 2, 139 | 0.566 | 864 | 39.58 | 0.32 |
| 77 | -32.39 | -12.12 | 107 | 6 | 0.195 | 480 | 39.31 | 0.34 |
| 78 | -33.12 | 9.19 | | 151 | 0.239 | 222 | 38.19 | 0.30 |
| 79 | -34.03 | -35.91 | 108 | 40 | 0.246 | 299 | 38.90 | 0.32 |
| 80 | -34.82 | -45.10 | 106, 109 | 36, 79 | 0.282 | 401 | 39.13 | 0.32 |
| 81 | -39.93 | 71.19 | 112 | 20 | 0.574 | 350 | 39.06 | 0.29 |
| 82 | -39.98 | -17.22 | 113 | 84 | 0.240 | 302 | 38.64 | 0.32 |
| 83 | -43.27 | 63.54 | 117 | 65 | 0.548 | 392 | 38.74 | 0.34 |
| 84 | -47.68 | -56.37 | 115 | 11 | 0.365 | 512 | 39.22 | 0.34 |
| 85 | -50.75 | 139.24 | 121 | 44 | 1.003 | 585 | 38.89 | 0.30 |
| 86 | -54.49 | 57.58 | 123 | 41 | 0.576 | 468 | 38.89 | 0.32 |
| 87 | -59.47 | -66.74 | 127 | 53 | 0.442 | 520 | 38.80 | 0.34 |
| 88 | -59.47 | -84.31 | 125 | 123 | 0.512 | 314 | 38.37 | 0.33 |
| 89 | -60.27 | -28.49 | | 102 | 0.362 | 422 | 38.51 | 0.30 |
| 90 | -62.87 | 50.73 | | 30 | 0.589 | 500 | 38.95 | 0.33 |
| 91 | -69.73 | 96.68 | 132 | 119 | 0.853 | 484 | 38.41 | 0.30 |
| 92 | -72.34 | 5.68 | | 59 | 0.478 | 678 | 38.77 | 0.31 |
| 93 | -74.09 | 38.26 | | 83 | 0.600 | 422 | 38.65 | 0.31 |
| 94 | -88.32 | -48.50 | 134 | 70 | 0.535 | 412 | 38.73 | 0.32 |
| 95 | -89.34 | 22.39 | 135 | 68 | 0.635 | 643 | 38.73 | 0.33 |
| 96 | -93.64 | -0.27 | 139 | 98, 147 | 0.602 | 445 | 38.73 | 0.29 |
| 97 | -97.78 | -121.71 | 138 | 27 | 0.772 | 439 | 38.97 | 0.33 |
| 98 | -125.77 | 13.38 | | 184 | 0.840 | 233 | 38.00 | 0.37 |

4. *Results and discussion.* According to our observations only 98 HII regions have detectable Hα and [NII] emissions. This number of identified HII regions respectively is about 1.4 and 2.9 times less that reported by Hodge and Kennicutt [5] and Feinstein [6]. This difference, at first, caused by the fact that we select only HII regions which radiate in both Hα and [NII] lines, at second, Hodge and Kennicutt [5] and Feinstein [6] used higher resolution and sensitivity Hα observational material. Similar to the general population of spiral galaxies the outermost HII region in the sample is situated approximately at R(0), while the mean distance is on 0.47R(0) (e.g. Athanassoula et al. [8]).

4.1. *Distribution of the diameters of HII regions.* The histogram of the linear effective diameter for 98 HII regions of NGC6384 is shown in Fig.2. The maximum of $N(D)$ occurs in the range 200 – 500 pc. The short left cut of the distribution is due to the fact that we do not pick up the HII regions smaller than 1.7″ x 1.7″ (about 180 pc x 180 pc). In Fig.3 is plotted the number of HII regions with linear diameter larger than $D$ as a function of diameter $D$ in parsecs. The frequency distribution of the HII regions diameters can be approximated by the exponential low (van den Bergh [18]), i.e. $N = N_o \exp(- D/D_o)$ with $N_o = 312$ and $D_o = 217$ pc. $D_o$ has the value predicted for a galaxy of its measured absolute luminosity according to the observational plot of $D_o$ versus luminosity in Hodge [19]. Fig.4 shows the relation of the relative distances of HII regions (R/R(0)) versus their linear diameters. The linear least-square fit is shown by solid line which has following form and coefficient of correlation:

$D$(pc) = (236 ± 77)R/R(0) + (305 ± 40);     r = 0.299 ± 0.097

Fig.4 demonstrates that there is a shallow positive correlation between these two parameters, with the tendency of larger HII regions to be located in outer regions of the galaxy (Hodge [19]).

4.2. *Luminosity function of the HII regions.* Luminosity function of HII regions of NGC6384 previously was studied by Kennicutt et al. [20] and Feinstein [6]. In these studies the contribution of [NII] lines luminosities in observed Hα luminosity of individual HII regions was not accounted (Feinstein [6]), or was accounted adopting average [NII]/Hα ratio for the galaxies similar to NGC6384 morphology (Kennicutt [21], Kennicutt et al. [20]). In our study Hα luminosities for individual HII regions are determined more correctly hence repetition of their luminosity function study is valuable. The luminosity function of HII regions is usually represented by the power low $dN(L) \sim L^{\alpha}\, dL$ between $[L_{min}, L_{max}]$ and zero elsewhere. Here $dN(L)$ is the number of HII regions within the luminosity range $[L, L+\Delta L]$. Fig.5 shows the log-log cumulative luminosity function for NGC6384 HII regions. The points represent the logarithms of cumulative numbers of HII regions per fixed luminosity interval 1.0E+38 ergs/sec. When the statistics are insufficient $N(L)$ is calculated from a larger interval. Constructed luminosity function has relatively shallow slope at low luminosities ($\alpha$ = - 0.4), an abrupt turnover at $\log L$(Hα) ≈ 38.75, and sharper slope at higher luminosities ($\alpha$ = - 2.3). According to Kennicutt et al. [20] this type of double power laws profile (type II LF) is common property of Sb – Sbc type spirals. The index ($\alpha$ = - 2.3) of the power law in the higher luminosity segment of LF is shallower than the average value ($\alpha$ = - 3.3, but see Rozas et al. [22] reported by Kennicutt et al. [20]), which can be caused by the low resolution of our observations according to which several close by HII regions merge forming larger complexes. The index ($\alpha$ = - 0.4) of the power law in the low luminosity segment is shallower than reported ($\alpha$ = - 1.3) by Kennicutt et al. [20]), which partially can be stipulated by incompleteness of HII regions sample at the faint end. Important conclusion of luminosity function study is the existence of abrupt turnover at $\log L$(Hα) ≈ 38.75 and its double power laws profile. Discovery of type II luminosity function for NGC6384 with similar data of Kennicutt et al. [20], Feinstein [6], Rozas et al. [22] proves that existence of type II galaxies is real and can be explained that there is a physical mechanism which inhibits the formation of giant and supergiant HII complexes in early type spirals (Kennicutt et al. [20]), or type II galaxies are undergoing or have recently undergone an increase in their star formation rates. Hence, the observed luminosity function comes from the usual star formation rates plus a recent burst, which makes the faint side of the luminosity function shallower and adds giant and supergiant HII regions in the bright side (Feinstein [6]) This discovery also is important, since it shows that existence of type II luminosity function does not depend from the galaxy nuclear activity level.

4.3. *Luminosity and diameter correlation.* The form of the correlation between luminosity and diameter is important parameter to determine inner conditions in the HII regions. For the simplest case of constant density, ionization-bounded HII region the cube of the Strömgren diameter should scale the ionizing luminosity. For our HII regions Fig.6 shows the correlation between logarithms of luminosities and diameters. Large scatter is partly due to measuring uncertainties, but also it possibly reflects the large variation in the internal structure of HII regions (Kennicutt [23]). In the Fig.6 the linear least-square fit is shown by solid line which has following form and coefficient of correlation:

$$\log L(\mathrm{H}\alpha) = (1.665 \pm 0.170)\log D(\mathrm{pc}) + (34.465 \pm 0.440); \quad r = 0.707 \pm 0.072$$

The slope of this relation is close to the average slope of the above mentioned Strömgren luminosity diameter relation, which is indicated by the arrow in Fig.6. Such a correlation was discovered for HII regions population of many galaxies (e.g. Kennicutt [21], Rozas et al. [22]) and it works for the galaxies with any level of nuclear activity.

4.4. *[NII]/Hα ratio distribution and other correlations.* Main tools to study the macroscopic properties of the disk galaxies which possibly drive their chemical evolution are HII regions. Heavy-element, particularly oxygen, abundances of HII regions is one of the mine parameters which often were in use (e.g. Zaritsky et al. [24], Melbourne and Salzer [25]). In comparison with oxygen the nitrogen abundance study is less frequent but valuable, since it helps to solve the problem of ratio between primary and secondary production of the heave-elements in the galaxies (e.g. Kennicutt et al. [26]). In this study we selected 98 HII regions which have detectable [NII]λ6584 emission. Nitrogen abundance in HII regions is strongly proportional to the [NII]λ6584/Hα line ratio and hence this ratio can be use as a marker for its study (e.g. Burenkov [27]). Fig.7 shows [NII]/Hα line ratio versus galactocentric distance. The linear least-square fit is shown by solid line which has following form and coefficient of correlation:

$$[\mathrm{NII}]/\mathrm{H}\alpha = (-0.038 \pm 0.011)R/R(0) + (0.323 \pm 0.006); \quad r = -0.338 \pm 0.096$$

Fig.7 clearly demonstrates that there is a negative radial gradient of [NII]/Hα lines ratio, it means, nitrogen abundance in NGC6384. In the central region of the galaxy nitrogen abundance is higher than in the periphery. The value of reported radial gradient (- 0.038 ± 0.011) is in good agreement with the similar value (- 0.041) obtained by Pilyugin et al. [13] using nitrogen abundances data for 9 HII regions.

4.5. *1971L supernova in the galaxy.* Currently more data have been accumulated which show that significant fraction of type Ia SN events in late spirals and irregular galaxies originates in relatively young stellar component (e.g. Mannucci et al. [28], Petrosian et al. [29]). SN 1971L (RA = $17^h 32^m 26^s.1$, Dec = + 07° 04′ 02″; J2000.0) discovered in NGC6384 is well studied. Particularly its location in respect to nearest spiral arm and HII region were studied by Van Dyk [30] and Bartunov et al. [31] and results of HST study of its accurate location was reported by Boffi et al. [32]. According to Boffi et al. [32] a blue arc-shaped patch is present just close to the position of SN. HST images resolve the underlying spiral structure and several compact sources of emission are present within the uncertainty ring. Connection of the SN with spiral structure also was stressed by Bartunov et al. [31]. SN 1971L does not show obvious connection to the nearest HII regions. According to Van Dyk [30] and Bartunov et al. [31] closest to SN HII region has an offset 32E 15N and lies at the distance of 16.8″. This HII region is not catalogued by Hodge and Kennicutt [5] and Feinstein [6], but well seen in the Hα map of the galaxy presented by Knapen et al. [7]. According to our Hα image we reexamine the location of SN in the galaxy. At first, the location of SN in the spiral arm is confirmed. Its location relatively to closest HII region was studied using two data sets for SN position, at first, the offset 27E 20N reported by the Dunlap [33], and at second the SN coordinates presented in the NED. If the offset 27E 20N is in used (whose accuracy is about ± 10″) above mentioned HII region (Van Dyk [30]) is the closest one. If NED coordinates are in use, closest to the SN position is #53 (F81) HII region (offset 21.96E 26.87N) which lies at the distance of 8.6″. The location of the SN in spiral arm and close by distribution to the HII region possibly indicates that its progenitor cannot belong to an old, evolved population.

5. *Conclusions.* The main results of the present paper, where we analyzed the statistics and properties of HII regions in the spiral galaxy NGC6384 are following:

1. Using narrow band interferometric Hα and [NII]λ6584 images of the galaxy we identified 98 HII regions radiating in both emission lines. Appropriate catalogue which includes the positions of the discovered HII regions, their identification as Hodge and Kennicutt [5] and Feinstein [6] objects, their relative distances, effective diameters, Hα luminosities and [NII]/Hα ratios is presented.
2. The integrated distribution function of the HII region linear diameters can be well fitted by the exponential function. The characteristic diameter has the value ($D_o$ = 217 pc) predicted for a galaxy of its measured absolute luminosity and lies between the range reported previously in the literature for the galaxies of similar morphology.
3. The luminosity function of HII regions has double power laws profile (type II LF), with relatively shallow slope at low luminosities ($\alpha$ = - 0.4), an abrupt turnover at log $L$(Hα) ≈ 38.75, and sharper slope at higher luminosities ($\alpha$ = - 2.3). This type of profile is common property of Sb – Sbc type spirals.
4. Despite of the large scatter of the data the form of the correlation between luminosity and diameter of HII regions confirms that in general they are constant density radiation-bound systems.
5. According to the [NII]/Hα ratio data for HII regions there is a negative radial gradient of [NII]/Hα, it means, nitrogen abundance in NGC6384. In the central region of the galaxy nitrogen abundance is higher than in the periphery. The value of reported radial gradient (- 0.038 ± 0.011) is in good agreement with the similar value (- 0.041) presented in the literature.
6. The properties of the HII regions population of this AGN galaxy does not differs significantly from the properties of the HII regions population of the “normal” galaxies. This probably indicates that the nuclear activity of the galaxies is local process.

7. Reexamining the location of the type Ia SN 1971L in the galaxy, we confirm that it lies on the spiral arm of NGC6384, about 8.6″ far from the closest HII region #53 (F81). Such a location can be taken as a prove that the progenitor of this SN do not belong to an old, evolved stellar population.

[1] V.A.Ambartsumian Byurakan Astrophysical Observatory,
Armenia, e-mail: hakartur@rambler.ru
[2] Observatoire de Marseille, 2 Place Le Verrier, F-13248 Marseille Cedex 04,
France, e-mail: Jacques.Boulesteix@oamp.fr

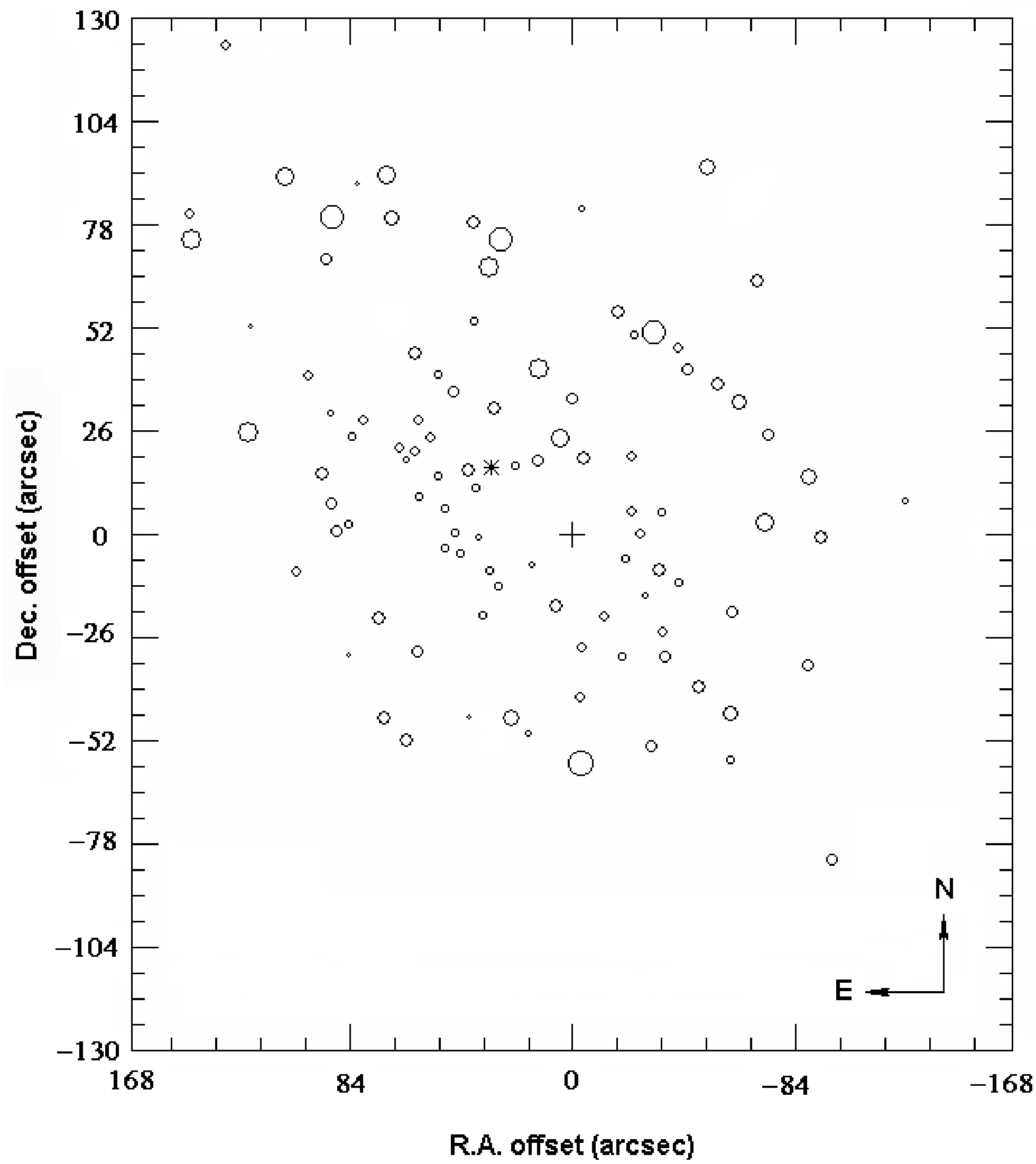

Fig.1. An identification chart of the measured HII regions. The size of the circles is proportional to the diameters of HII regions. The right ascension a declination offsets are relative to the nuclear position (marked by a plus sign) as described in the text. The position of SN 1971L is marked by asterisk (*).

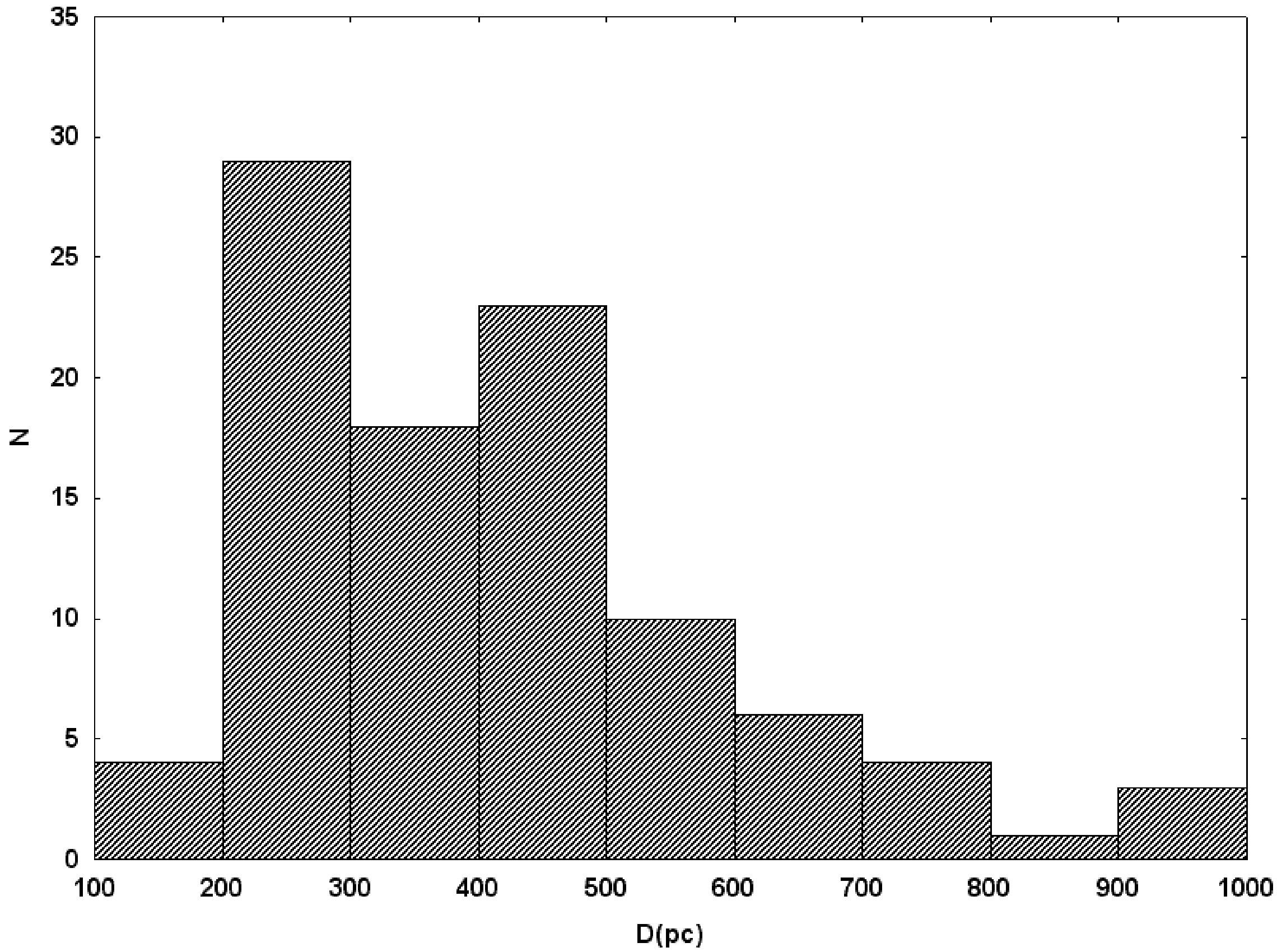

Fig.2. Histogram of the effective linear diameters of HII regions.

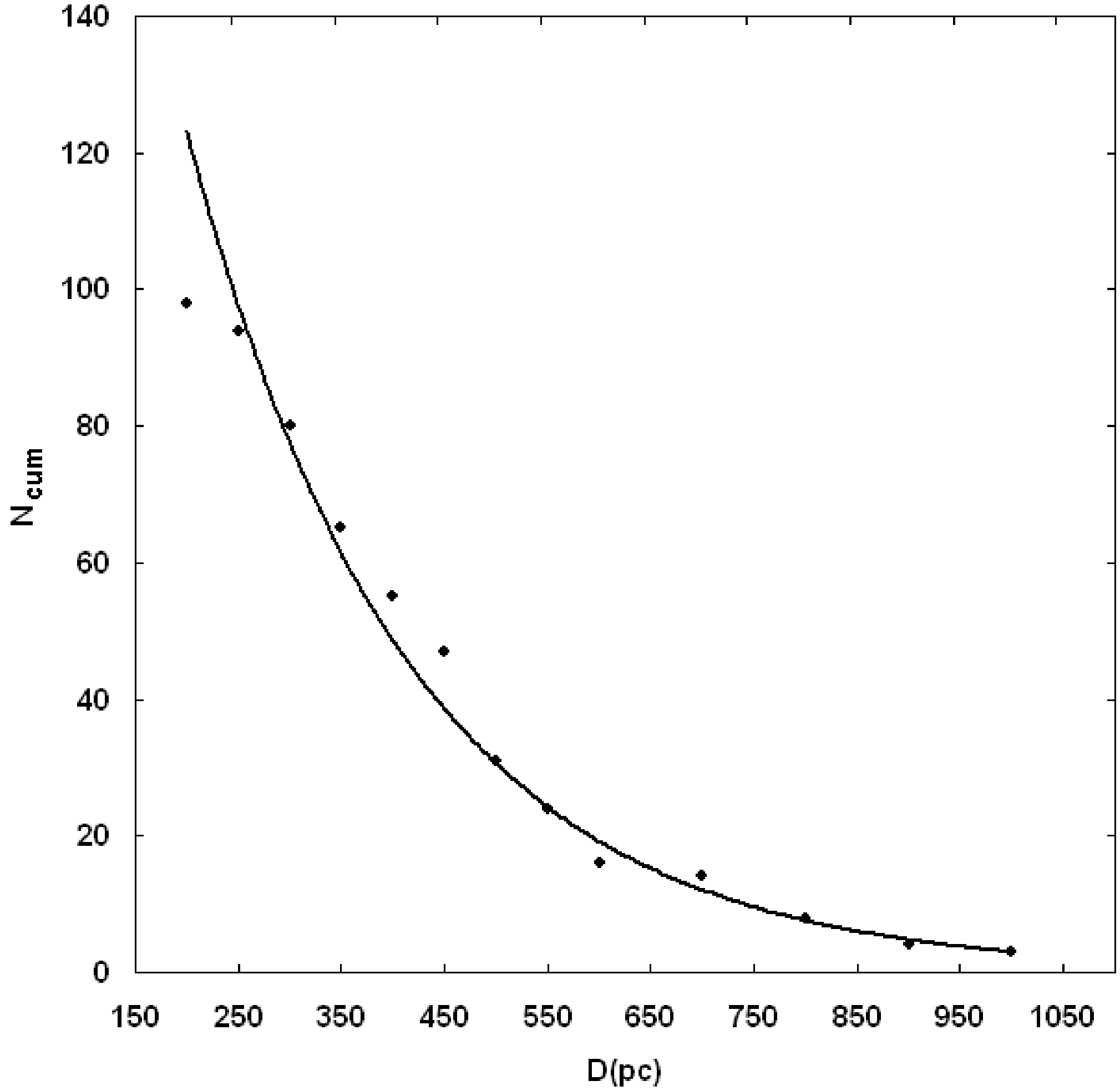

Fig.3. Cumulative diameter function of HII regions. The best exponential fit is drawn.

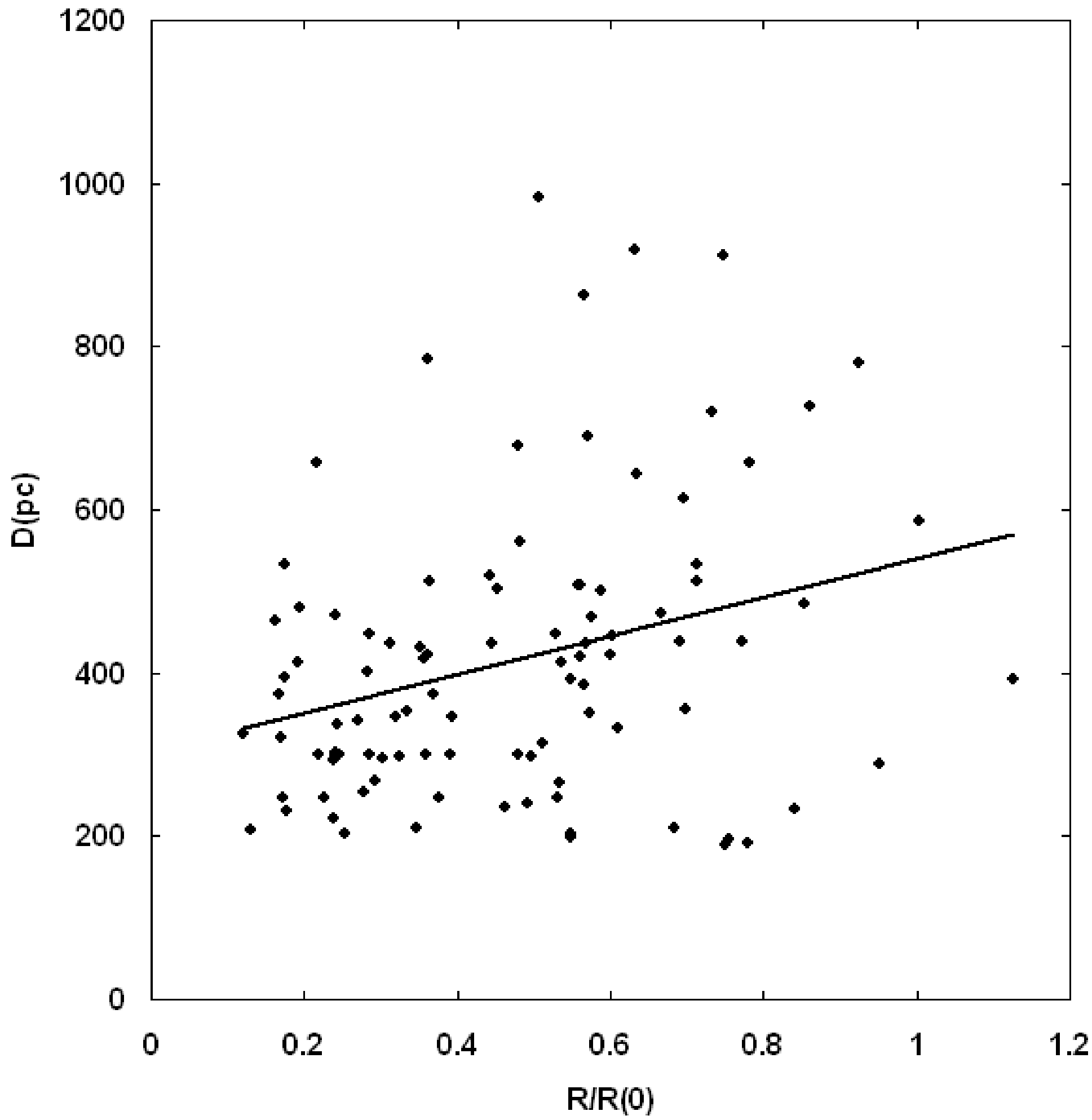

Fig.4. The relation of the relative distances of HII regions (R/R(0)) versus their linear diameters. The straight line indicates best linear fit.

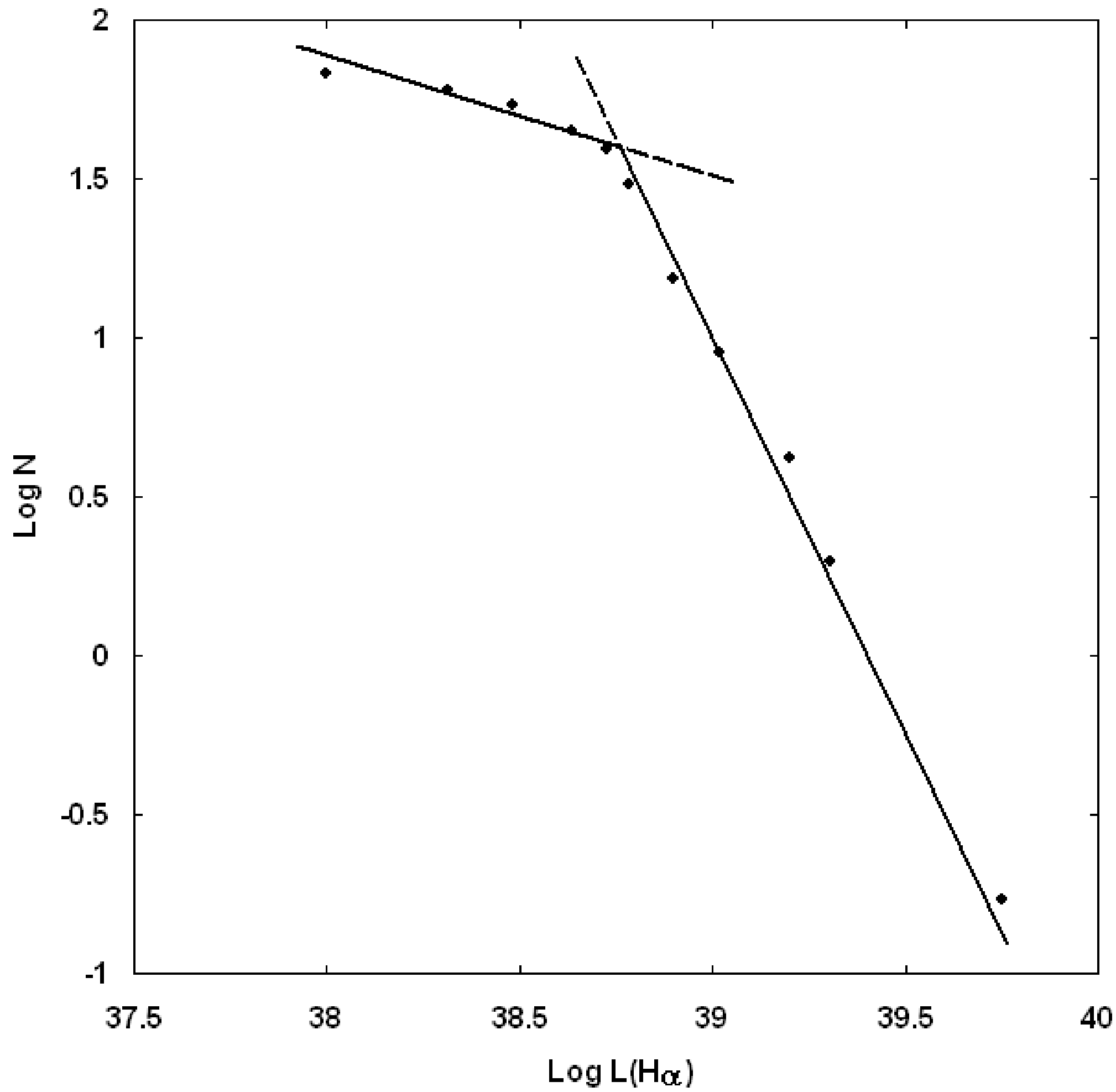

Fig.5. Hα luminosity function of HII regions. Bi-linear fit for the luminosity function is shown. The change in slope is seen at log $L(\mathrm{H}\alpha) \approx 38.75$ ergs/sec.

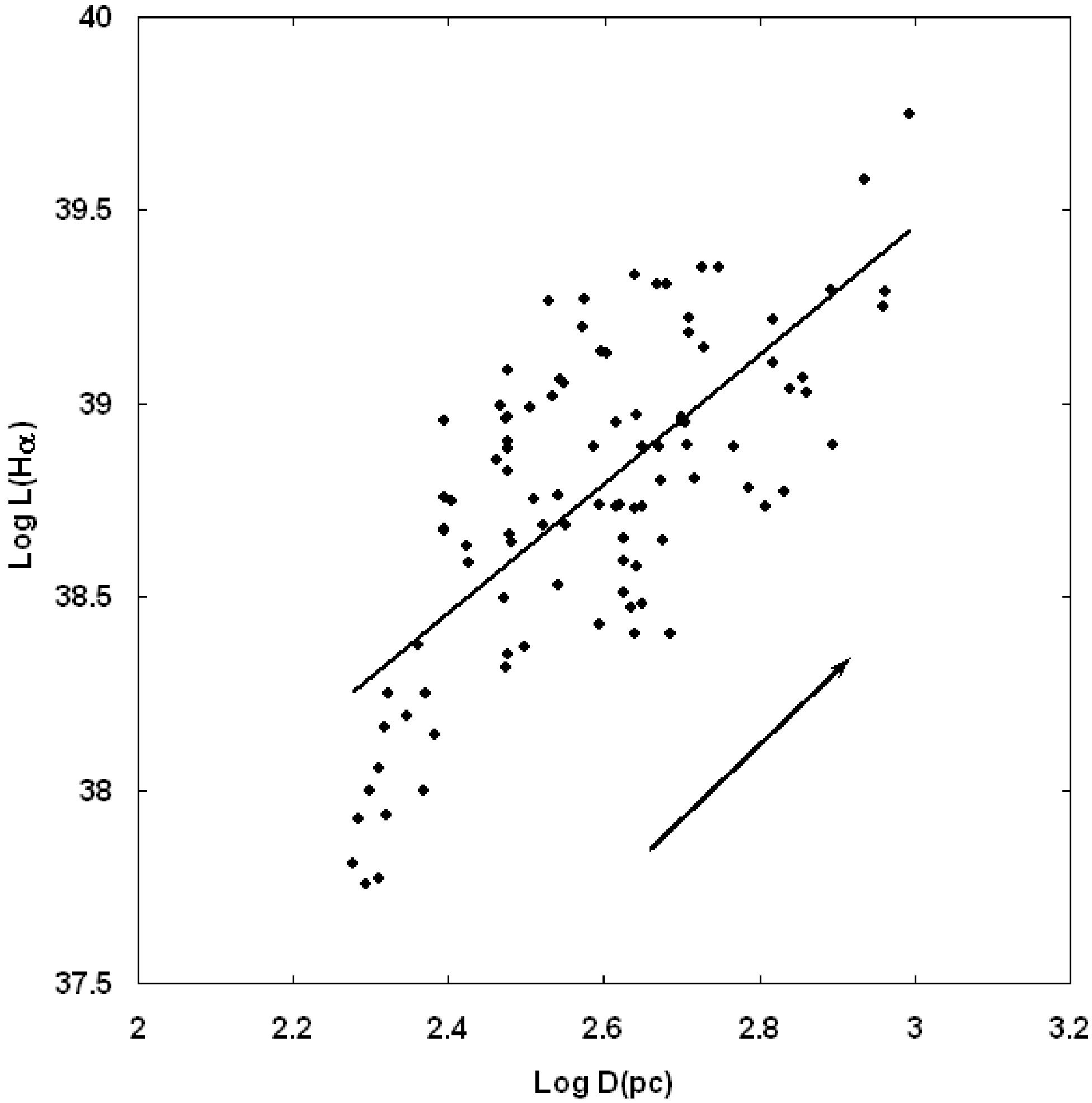

Fig.6. Logarithm of the Hα luminosity versus logarithm of linear diameter of HII regions. The straight line indicates best linear fit. The arrow indicates the slope expected for constant density radiation-bounded nebulae.

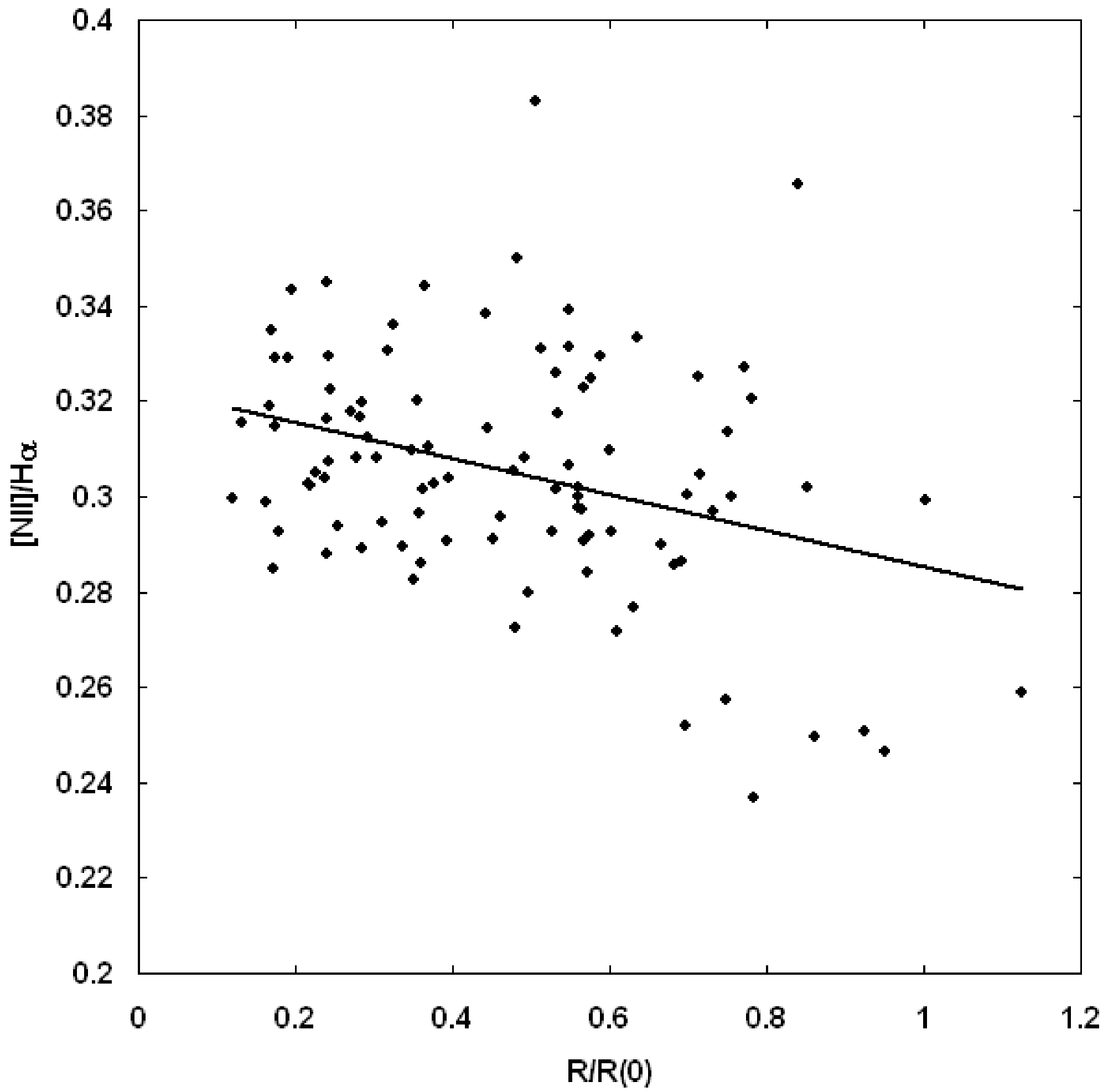

Fig.7. [NII]/Hα line ratio versus relative galactocentric distance. The straight line indicates best linear fit.